\documentclass[10pt,twocolum,final]{IEEEtran}
\usepackage[utf8]{inputenc}
\usepackage{bm}
\usepackage{amsmath}
\usepackage{amssymb}
\usepackage{graphicx}

\usepackage[
            linkcolor=black, 
            anchorcolor=black, 
            citecolor=black, 
            urlcolor=black
           ]{hyperref}

\makeatletter

\usepackage{amsfonts}
\usepackage{color}
\usepackage{multicol}
\usepackage{subfigure}
\usepackage{bm}
\usepackage{latexsym}
\usepackage{stfloats}
\usepackage{float}
\usepackage{exscale}
\usepackage{relsize}
\usepackage{cite}
\usepackage{cases}
\usepackage{algorithm}
\usepackage{algpseudocode}
\usepackage{graphics}
\usepackage{epsfig}
\usepackage{epstopdf}
\usepackage{breqn}
\usepackage{enumerate}
\usepackage{multirow}
\usepackage{caption}
\usepackage{xcolor}
 \usepackage{array}

\makeatother

\vspace{-0.5cm}
\vspace{-1.5cm}
\begin{document}
\title{Transmission Design for Active RIS-Aided Simultaneous Wireless
Information and Power Transfer\vspace{-0.3em}}
\IEEEoverridecommandlockouts
\author{
        Hong Ren,~\IEEEmembership{Member,~IEEE},
        Zhiwei Chen,
        Guosheng Hu,
        Zhangjie~Peng,

        Cunhua Pan,~\IEEEmembership{Member,~IEEE},
        and Jiangzhou Wang, ~\IEEEmembership{Fellow,~IEEE}

\vspace{-0.5em}
\vspace{-0.5cm}

\thanks{This work was supported in part by the National Natural Science Foundation of China (62201137),
in part by the
Natural Science Foundation
of Shanghai under Grant 22ZR1445600, in part by the National Natural
Science Foundation of China under Grant 61701307, in part by the open
research fund of National Mobile Communications Research Laboratory,
Southeast University under Grant 2018D14, and in part by the National
Natural Science Foundation of China (62101128) and Basic Research Project
of Jiangsu Provincial Department of Science and Technology (BK20210205).\emph{ (Corresponding authors: Zhangjie Peng and Zhiwei Chen.)}}
\thanks{Hong Ren and Cunhua Pan are with the National Mobile Communications Research Laboratory,
Southeast University, Nanjing 210096, China. (e-mail: hren@seu.edu.cn; cpan@seu.edu.cn).}
\thanks{Zhiwei Chen is with the College of Information, Mechanical and Electrical Engineering,
 Shanghai Normal University, Shanghai 200234, China (e-mail: 1000497437@smail.shnu.edu.cn).}
 \thanks{Guosheng Hu is with the Shanghai Technical Institute of Electronics \&
Information, Shanghai 201411, China (e-mail: huguosheng@stiei.edu.cn).}
\thanks{Zhangjie Peng is with the College of Information, Mechanical, and Electrical Engineering, Shanghai Normal University, Shanghai 200234, China, also with the National Mobile Communications Research Laboratory, Southeast University, Nanjing 210096, China, and also with the Shanghai Engineering Research Center of Intelligent Education and Bigdata, Shanghai Normal University, Shanghai 200234, China (e-mail: pengzhangjie@shnu.edu.cn).}
 \thanks{Jiangzhou Wang is with the School of Engineering, University of Kent, CT2 7NT
Canterbury, U.K. (e-mail: j.z.wang@kent.ac.uk).}
}


\maketitle

\newtheorem{lemma}{Lemma}
\newtheorem{theorem}{Theorem}
\newtheorem{remark}{Remark}
\newtheorem{corollary}{Corollary}
\newtheorem{proposition}{Proposition}
\newcounter{TempEqCnt}
\vspace{-1cm}
\begin{abstract}
Reconfigurable intelligent surface (RIS)
is a revolutionary technology to enhance both the spectral efficiency and energy efficiency of
 wireless communication systems.
However, most of the existing contributions mainly focused  on the study of passive RIS, which suffers from the ``double fading'' effect.
On the other hand, active RIS,  which is equipped with amplifiers, can effectively address this issue.
In this paper, we propose an active RIS-aided  
simultaneous wireless information and power transfer (SWIPT) system.
Specifically, we maximize the weighted sum rate of the information receivers, 
subject to the minimum power received at all  energy receivers, amplification power constraint at the active RIS, and the maximum transmit power constraint
at the base station (BS).
By adopting alternating optimization framework,
 suboptimal solutions are obtained.
Simulation results show that the active RIS-aided SWIPT system has higher performance gain with the same power budget.


\end{abstract}

\vspace{-0.75cm}
\section{Introduction}
\vspace{-0.2cm}
Reconfigurable intelligent surface (RIS), composed of a large number of reflecting elements,
has received extensive research attention in both academia and industry\cite{9318531,9847080,8796365,9090356}.
Specifically,
RIS can dynamically adjust the electromagnetic properties of the reflecting elements in a programmable way,
and then reconfigure the wireless propagation environment in a desired way\cite{9180053,9366346}.

However,
most of the existing contributions mainly focused on the passive
RIS-aided communication systems,
which suffer from the ``double fading'' effect\cite{9734027}.
To address this issue,
active RIS has been
proposed in \cite{zhang2021active,9377648},
which is equipped with some amplifiers.
{Different from multiple input multiple
output (MIMO) relay with high power cost and additional
time/frequency resource, e.g., amplify and forward (AF) relay, the active RIS
basically inherits the hardware structure of passive RIS,
while equipping with
 a set of low-power reflection-type amplifiers.}
As a result,
active RIS can not only tune the phase of the reflected signals,
but also amplify the power of the reflected signals.
Recently, the authors of \cite{9734027} have rigorously demonstrated
that the active RIS-aided single-input single-output (SISO) system
is superior to the passive RIS in terms of the achievable date rate
when both systems have the same power budget.
On the other hand,
simultaneous wireless information and power transfer (SWIPT) is envisioned as a
promising technology in future internet of things (IoT)\cite{9110849,9494365,9783020}.
The authors of \cite{9110849} investigated a passive  RIS-aided SWIPT system and showed that
the passive RIS can enhance the data rates at the information receivers (IRs),
while ensuring the minimum power requirements of the received power at the energy receivers (ERs).
{In \cite{9494365}, the authors studied an optimization problem of maximizing the minimum rate of
the IRs in the passive RIS-aided SWIPT system with imperfect channel state information (CSI).
And the authors of \cite{9783020} aimed to maximize the data rate by
proposing a joint time-switching and phase-shifting solution for passive RIS-assisted SWIPT communications.
However,
the above literatures \cite{9110849,9494365,9783020} about passive RIS-aided SWIPT systems suffer from the ``double fading'',
and for the SWIPT systems with requirements for both energy reception and information reception,
there are no literature investigating whether the active RIS with low-power amplifiers performs better than the MIMO relay with complex hardware structure.}

{Against the above background,
we consider an active RIS-aided SWIPT downlink system,
and aim to maximize the downlink weighted sum rate (WSR).
Different from the passive RIS \cite{9110849,9494365,9783020},
it is noted that the active RIS
introduces a new optimization variable due to its ability to amplify the reflected signal,
generates the non-ignorable thermal noise,
and adds an output signal power constraint of the active RIS,
which makes the optimization problem more challenging.
To solve the non-convex problem, our contributions of this work are summarized as follows}

$1)$
By considering
 the active RIS-aided SWIPT system, we aim to  maximize the downlink weighted sum rate (WSR) of the IRs, by
 jointly optimizing the transmit beamforming at the base station (BS) and the
reflecting coefficients at the active RIS, subject to the minimum power
harvested at all ERs, amplification power constraint at the active
RIS, and the maximum transmit power constraint at the BS.

$2)$ {By adopting  alternating optimization (AO) framework,
we transform the objective function by fractional programming (FP) method,
and utilize the first-order Taylor approximation to linearize the non-convex constraint of the active RIS  amplification power.
Then, we effectively solve the subproblems and obtain the suboptimal solutions.}

$3)$ {Simulation results show that the active RIS can achieve higher downlink WSR
than the passive RIS/AF relay in an SWIPT system \cite{9110849,9494365,9783020,7907183} with the same power budget.
And as the location of the RIS is closer to the locations of the ERs, it can achieve higher downlink WSR.
In addition, the appropriate number of RIS reflecting elements is enough to enable the SWIPT system to achieve good performance.}

\vspace{-0.2cm}
\section{System Model}
\vspace{-0.1cm}
As shown in Fig. 1,
we consider an active RIS-aided multiuser multiple input single output (MISO) downlink system,
where an RIS with $L$ reflecting elements is deployed to assist SWIPT.
The system is composed of
a BS with $N$ antennas,
$K_{I}$  single-antenna IRs,
and $K_{E}$ single-antenna ERs.

The signal transmitted from the BS is expressed as
\vspace{-0.2cm}
\setcounter{equation}{0}
\begin{align}\label{0}
{\bf t} & =\sum\limits_{k=1}^{K_{I}}{\bf w}_{k}{s}^{{IR}}_{k}+{\bf v},
\end{align}
where 
${\bf w}_{k}\in{\mathbb{C}}^{N\times 1}$ is the beamforming vector for the $k$-th IR,
${s}^{ {IR}}_{k}\sim\mathcal{CN} (0, 1), k\in\{1,\cdot\cdot\cdot, K_{I}\}$ is the transmit information symbol  for the
$k$-th IR,
and ${\bf v}\in{\mathbb{C}}^{N\times 1}  \sim\mathcal{C}\mathcal{N}(\mathbf{0},{\mathbf{V}})          $ is the  energy signal vector,
where ${\mathbf{V}}$ is the covariance matrix of the energy signal vector.

The channels spanning from the BS to the active RIS, from the BS
to the $k$-th IR, from the BS to the $i$-th ER, from the RIS
to the $k$-th IR, and from the RIS to the $i$-th ER are denoted as  ${\mathbf{Q}}\in{\mathbb{C}}^{L\times N}$, ${\bf{g}}_{ d,k}\in{\mathbb{C}}^{N\times 1}$,
${\bf{h}}_{ d,i}\in{\mathbb{C}}^{N\times 1}$, ${\bf{g}}_{ r,k}\in{\mathbb{C}}^{L\times 1}$, and ${\bf{h}}_{ r,i}\in{\mathbb{C}}^{L\times 1}$, respectively.

The reflected and amplified signal at the active RIS can be modeled as follows
\vspace{-0.3cm}
\begin{align}
{\bf t}_{\mathrm r}  ={\bm{\Phi}} \left({\mathbf{Q}}{\bf t}+{\bf n}_{\mathrm {RIS}}\right),
\end{align}
where $\bm{\Phi}={\textrm {diag}}\left({a_{1}e^{j\phi_{1}},\cdots,a_{l}e^{j\phi_{l}},\cdots,a_{L}e^{j\phi_{L}}}\right)$ denotes the reflection matrix of the active RIS,
$\phi_{l}$ and $a_{l}$ are the phase shift and the amplitude of the $l$-th element, respectively.
The thermal noise generated by the active RIS  cannot be
neglected,
and ${\bf n}_{\mathrm {RIS}}\sim\mathcal{C}\mathcal{N}(\bm{0},\delta_{r}^{2}{\mathbf{I}})$ denotes the thermal noise of the active RIS.

\begin{figure}
\vspace{-1.6cm}
\includegraphics[scale=0.6]{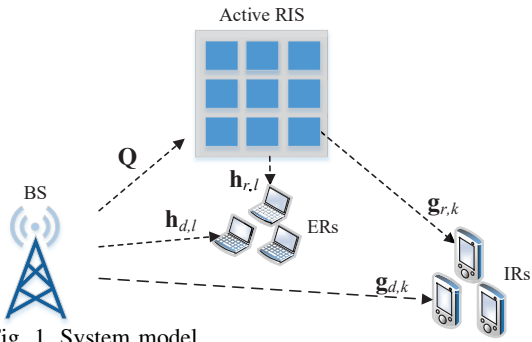}
\captionsetup{font={small},justification=raggedright}
\vspace{-0.5cm}
\caption{ System model. }
\label{Figsysmodel}
\vspace{-0.35cm}
\end{figure}

The received signal at the $k$-th IR can be written as
\begin{flalign}
&{ y}^{{ {IR}},k} \!={\bf{g}}_{ d,k}^{\mathrm H}{\bf t}+{\bf{g}}_{ r,k}^{\mathrm H}{\bf t}_{\mathrm r}    +{ n}_{ \mathrm{IR}}&\nonumber\\
&\quad\quad ={\bf{g}}_{k}^{\mathrm H}{\bf t}  +{\bf{g}}_{ r,k}^{\mathrm H}{\bm{\Phi}} {\bf n}_{\mathrm {RIS}} +{ n}_{ \mathrm{IR}}\nonumber\\
&\quad\quad={\bf{g}}_{k}^{\mathrm H}{\bf w}_{k}{s}^{{IR}}_{k}\!\!+\!\!\!\!\!\!\sum\limits_{i=1,i\neq k}^{K_{I}}\!\!\!\!\!{\bf{g}}_{k}^{\mathrm H}{\bf w}_{i}{s}^{{IR}}_{i}\!+\!{\bf{g}}_{k}^{\mathrm H}{\bf v}\!+{\bf{g}}_{ r,k}^{\mathrm H}{\bm{\Phi}} {\bf n}_{\mathrm {RIS}} +{ n}_{ \mathrm{IR}},
\end{flalign}
where ${\bf g}_{k}^{\mathrm H}\triangleq{\bf g}_{ d,k}^{\mathrm H}+ {\bf{g}}_{ r,k}^{\mathrm H}{\bm{\Phi}}{\mathbf{Q}}$, and ${ n}_{ \mathrm{IR}}\sim\mathcal{C}\mathcal{N}(0,\delta_{\mathrm{IR}}^{2})$ is the additive white Gaussian
noise (AWGN).
{Unlike the passive RIS model\cite{9110849,9494365,9783020},
the active RIS consisting of amplifiers has the ability to amplify the power of the reflected signal.
Thus, the reflection matrix $\bm{\Phi}$ has non-unity amplitude components}.

Then, the signal-to-interference-plus-noise ratio (SINR) of the $k$-th IR is expressed as
\vspace{-0.5cm}
\begin{align}
\gamma_{k}&=    \frac{|{\bf{g}}_{k}^{\mathrm H}{\bf w}_{k}|^{2}} { \sum\limits_{i=1,i\neq k}^{K_{I}}|{\bf{g}}_{k}^{\mathrm H}{\bf w}_{i}|^{2}+
{\bf{g}}_{k}^{\mathrm H}{\mathbf{V}}{\bf{g}}_{k}+
 \delta_{r}^{2} \|{\bf g}_{ r,k}^{\mathrm H}{\bm{\Phi}} \|^{2}   +\delta^{2}_{ \mathrm{IR}}  }.
\end{align}

Thus, the rate at the $k$-th IR is expressed as
\vspace{-0.3cm}
\begin{align}\label{26}
R_{k}={\mathrm{log_{2}}}\left(1+  \gamma_{k}\right).
\end{align}

\vspace{-0.3cm}
The received signal at the $i$-th ER can be written as
\vspace{-0.2cm}
\begin{align}
{y}^{{ {ER}},i} & ={\bf{h}}_{ d,i}^{\mathrm H}{\bf t}+{\bf{h}}_{ r,i}^{\mathrm H}{\bf t}_{\mathrm r}    +{ n}_{\mathrm {ER}}\nonumber\\
&={\bf{h}}_{i}^{\mathrm H}{\bf t}  +{\bf{h}}_{ r,i}^{\mathrm H}{\bm{\Phi}} {\bf n}_{\mathrm {RIS}} +{n}_{ \mathrm{ER}}\nonumber\\
&=\sum\limits_{k=1}^{K_{I}}{\bf{h}}_{i}^{\mathrm H}{\bf w}_{k}{s}^{{IR}}_{k}+{\bf{h}}_{i}^{\mathrm H}{\bf v}+{\bf{h}}_{ r,i}^{\mathrm H}{\bm{\Phi}} {\bf n}_{\mathrm {RIS}} +{n}_{\mathrm {ER}},
\end{align}
where ${\bf h}_{i}^{\mathrm H}\triangleq{\bf h}_{ d,i}^{\mathrm H}+ {\bf{h}}_{ r,i}^{\mathrm H}{\bm{\Phi}}{\mathbf{Q}}$, and ${ n}_{ \mathrm{ER}}\sim\mathcal{C}\mathcal{N}(0,\delta_{\mathrm{ER}}^{2})$ is the AWGN at the $i$-th ER.
Considering the fact that
both the data and energy signals transmitted by the BS are carried by the beamforming,
the harvested power at the $i$-th ER while ignoring the AWGN  power is given by
\vspace{-0.3cm}
\begin{align}
E_{i}=\eta_{i}\left(\sum\limits_{k=1}^{K_{I}}|{\bf{h}}_{i}^{\mathrm H}{\bf w}_{k}|^{2} +{\bf{h}}_{i}^{\mathrm H}{\mathbf{V}}{\bf{h}}_{i}+ \delta_{r}^{2}\|{\bf{h}}_{ r,i}^{\mathrm H}{\bm{\Phi}} \|^{2}\right),
\end{align}
where $\eta_{i}$ is the energy harvesting efficiency of the $i$-th ER.

\vspace{-0.5cm}
\section{Problem Formulation}

To satisfy the requirements of both  IRs and ERs,
we consider an optimization problem of maximizing the WSR of all IRs,
while satisfying the harvested power requirements of all ERs,
subject to the power constraints at the BS and active RIS.
Thus, the optimization problem is formulated as
\vspace{-0.2cm}
\begin{subequations}\label{question_1}
\begin{align}
&\max_{\{{\bf w}_{k}\},{\mathbf{V}},{\bm{\Phi}}}\quad   \sum\limits_{k=1}^{K_{I}}\alpha_{k}R_{k} &\\
&\quad\mbox{s.t.} \quad \quad  \|{\bm{\Phi}} {\mathbf{Q}}{\bf t}\|^{2}+\delta_{r}^{2}\|{\bm{\Phi}}\|^{2}\leqslant P_{\mathrm{RIS}}^{\mathrm{act}},\label{P_ris}\\
 & \quad\quad\quad\quad \|{\bf t}\|^{2}\leqslant P_{\mathrm {BS}}^{\mathrm{act}}\label{P_BS},\\
 & \quad\quad\quad\quad E_{ {i}}\geqslant P_{{i}}, i\in\{1,\cdots,K_{E}\},\label{P_ER}
\end{align}
\end{subequations}
where $\alpha_{k}$ is the weighting factor of the $k$-th IR,
 $P_{\mathrm{RIS}}^{\mathrm{act}}$ is the output signal power of the active RIS,
$ P_{\mathrm {BS}}^{\mathrm{act}}$ is the transmit power limit at the BS
and $P_{{i}}$ is the minimum harvested power threshold for the $i$-th ER.

Due to the fact that variables $\left\{ \{{\bf w}_{k}\},{\mathbf{V}},{\bm{\Phi}} \right\}$  are coupled together in the objective function of  Problem \eqref{question_1}, it is challenging to
solve Problem \eqref{question_1}.
We then exploit the fractional programming (FP) \cite{8314727} method to decouple the objective function of Problem \eqref{question_1},
and adopt the Alternate Optimization (AO) algorithm to obtain  the solutions
 in the next subsection.

\vspace{-0.5cm}
\subsection{FP method}
Firstly, we use the FP method to transform the objective function.
By introducing auxiliary variables $\bm{\tilde{\gamma}} = [\tilde{\gamma}_{1},\cdots,{\tilde{\gamma}_{K_{I}}}]^{\mathrm T}\in{\mathbb{C}}^{K_{I}\times 1}$,
the objective function of Problem \eqref{question_1} is equivalent to
\vspace{-0.3cm}
\begin{align}
&f_{a}({\bm{\tilde{\gamma}}},{\bf w}_{k},{\mathbf{V}},{\bm{\Phi}})= \sum\limits_{k=1}^{K_{I}} \alpha_{k} {\mathrm{log}}\left(1+{{\tilde{\gamma}}_{k}}\right)-\sum\limits_{k=1}^{K_{I}}\alpha_{k}{{\tilde{\gamma}}_{k}}\nonumber\\
&+\sum\limits_{k=1}^{K_{I}}\frac{\alpha_{k}{\left(1+{\tilde{\gamma}_{k}}\right)}|{\bf{g}}_{k}^{\mathrm H}{\bf w}_{k}|^{2}    }           { \sum\limits_{i=1}^{K_{I}}|{\bf{g}}_{k}^{\mathrm H}{\bf w}_{i}|^{2}+
{\bf{g}}_{k}^{\mathrm H}{\mathbf{V}}{\bf{g}}_{k}+
 \delta_{r}^{2} \|{\bf g}_{ r,k}^{\mathrm H}{\bm{\Phi}}\|^{2}   +\delta^{2}_{\mathrm{IR}} }.
\end{align}

We adopt the AO framework to obtain the optimal solutions.
For fixed variables $\{ \{{\bf w}_{k}\},{\mathbf{V}},{\bm{\Phi}} \}$,
by setting $\partial f_{a}/\partial{{\tilde{\gamma}_{k}}}$ to zero,
the optimal ${\tilde{\gamma}_{k}}^{\mathrm {opt}}$ is obtained as
\vspace{-0.2cm}
\begin{align}
{\tilde{\gamma}_{k}}^{\mathrm {opt}}={\gamma_{k}}\label{re_1}, k\in\{1,\cdots,K_{I}\}.
\end{align}

Then, we fix ${\tilde{\gamma}_{k}}$ and define a new function as
\vspace{-0.2cm}
\begin{align}
f_{b}&({\bm{\tilde{\gamma}}},{\bf w}_{k},{\mathbf{V}},{\bm{\Phi}}) \nonumber\\
&=\sum\limits_{k=1}^{K_{I}}\frac{\alpha_{k}{(1+{\tilde{\gamma}_{k}})}|{\bf{g}}_{k}^{\mathrm H}{\bf w}_{k}|^{2}    }           { \sum\limits_{i=1}^{K_{I}}|{\bf{g}}_{k}^{\mathrm H}{\bf w}_{i}|^{2}+
{\bf{g}}_{k}^{\mathrm H}{\mathbf{V}}{\bf{g}}_{k}+
 \delta_{r}^{2} \|{\bf g}_{ r,k}^{\mathrm H}{\bm{\Phi}}\|^{2}   +\delta^{2}_{\mathrm{IR}} }.
\end{align}

By introducing auxiliary variables ${\bm{\rho}}=[{\rho}_{1},\cdots,{{\rho}_{K_{I}}}]^{\mathrm T}\in{\mathbb{C}}^{K_{I}\times 1}$ and adopting the quadratic transform\cite{8314727},
we further recast $f_{b}$  as
\vspace{-0.4cm}
\begin{align}
&f_{c}({\bm{ \rho}},{\bm{\tilde{\gamma}}},{\bf w}_{k},{\mathbf{V}},{\bm{\Phi}})=2\sum\limits_{k=1}^{K_{I}}\sqrt{\alpha_{k}{(1+{\tilde{\gamma}_{k}})}}   {\mathcal{R} }\{ \rho_{k}^{*}{\bf{g}}_{k}^{\mathrm H}{\bf w}_{k}  \} \nonumber\\
&- \sum\limits_{k=1}^{K_{I}}|\rho_{k}|^{2}\left(\sum\limits_{i=1}^{K_{I}}|{\bf{g}}_{k}^{\mathrm H}{\bf w}_{i}|^{2}+
{\bf{g}}_{k}^{\mathrm H}{\mathbf{V}}{\bf{g}}_{k}+
  \delta_{r}^{2}\|{\bf g}_{ r,k}^{\mathrm H}{\bm{\Phi}} \|^{2}   +\delta^{2}_{ \mathrm{IR}} \right).
\end{align}

\vspace{-0.3cm}
Similarly, by setting $\partial f_{c}/\partial\rho_{k}$ to zero,
 we obtain the optimal $\rho_{k}^{opt}$   as
\vspace{-0.4cm}
\begin{align}
&\rho_{k}^{opt}=\frac{  \sqrt{\alpha_{k}{(1+{\tilde{\gamma}_{k}})}}  {\bf{g}}_{k}^{\mathrm H}{\bf w}_{k}               }       { \sum\limits_{i=1}^{K_{I}}|{\bf{g}}_{k}^{\mathrm H}{\bf w}_{i}|^{2}+
{\bf{g}}_{k}^{\mathrm H}{\mathbf{V}}{\bf{g}}_{k}+
 \delta_{r}^{2} \|{\bf g}_{ r,k}^{\mathrm H}{\bm{\Phi}} \|^{2}   +\delta^{2}_{ \mathrm{IR}}}, \nonumber\\
&\quad \quad \quad \quad \quad \quad k\in\{1,\cdots,K_{I}\}.\label{re_2}
\end{align}

\vspace{-0.3cm}
After obtaining the above optimal auxiliary variables,
in the next subsection,
we then focus on optimizing  $\{ \{{\bf w}_{k}\},{\mathbf{V}},{\bm{\Phi}} \}$,
given $\{{\bm \rho}$, ${\bm{\tilde{\gamma}}}\}$.

\vspace{-0.2cm}
\subsection{Optimizing ${\bf w}_{k}$ and ${\mathbf{V}}$ }
By defining ${\mathbf {W}}\triangleq[{\bf w}_{1}^{\mathrm T},\cdots,{\bf w}_{K_{I}}^{\mathrm T}]^{\mathrm T}$,
for fixed variables $\{{\bm \rho},{\bm{\tilde{\gamma}}},{\bm{\Phi}}\}$,  Problem \eqref{question_1} is expressed as
\vspace{-0.3cm}
\begin{subequations}\label{question_2}
\begin{align}
&\max_{{\mathbf {W}},{\mathbf {V}}}\quad   {\mathcal{R}} \{ {\bf b}^{\mathrm H}{\mathbf {W}}\}\!-\!  {\mathbf {W}}^{\mathrm H}\!{\mathbf {A}}_{1}\! {\mathbf {W}}\!- \! \sum\limits_{k=1}^{K_{I}}|\rho_{k}|^{2} \mathrm{{Tr}}\{ {\bf{g}}_{k}{\bf{g}}_{k}^{\mathrm H}\! {\mathbf{V}} \}                  \\
&\quad\mbox{s.t.} \quad \quad {\mathbf {W}}^{\mathrm H}{\mathbf {B}}{\mathbf {W}} +  \mathrm{{Tr}}\{{\mathbf{Q}}^{\mathrm H}{\bm{\Phi}}^{\mathrm H}{\bm{\Phi}}{\mathbf{Q}}{\mathbf{V}}\}  \leqslant \hat{P}_{\mathrm{RIS}}^{\mathrm{act}},\\
 & \quad\quad\quad\quad \|{\mathbf {W}}\|^{2}+\mathrm{{Tr}}\{{\mathbf {V}}\}\leqslant P_{\mathrm {BS}}^{\mathrm{act}},\\
 & \quad\quad\quad\quad {\mathbf {W}}^{\mathrm H}{\mathbf {D}}_{i}\!{\mathbf {W}}\!\!+\!\!\mathrm{{Tr}}\{\! {\bf{h}}_{i}{\bf{h}}_{i}^{\mathrm H} {\mathbf{V}} \!\}\!\!\geqslant\! P_{{i}}^{'}, i\!\in\!\{\!1,\!\cdots\!,K_{E}\!\},\label{P_ER_1}\\
  & \quad\quad\quad\quad {\mathbf {V}}\succeq 0,
\end{align}
\end{subequations}
where
\vspace{-0.3cm}
\begin{flalign}
&{\bf b}=[{\bf b}_{1}^{\mathrm T},{\bf b}_{2}^{\mathrm T},\cdots,{\bf b}_{{K_{I}}}^{\mathrm T}]^{\mathrm T},
{\bf b}_{k}^{\mathrm H}=2\sqrt{\alpha_{k}{(1+{\tilde{\gamma}_{k}})}}\rho_{k}^{*}{\bf{g}}_{k}^{\mathrm H},
\end{flalign}
\vspace{-0.8cm}
\begin{flalign}
&{\mathbf {A}}_{1}={\mathbf {I}}_{{K_{I}}}\otimes\sum\limits_{i=1}^{K_{I}}|\rho_{i}|^{2}{\bf{g}}_{i}{\bf{g}}_{i}^{\mathrm H},
\end{flalign}
\vspace{-0.5cm}
\begin{flalign}
{\mathbf {B}}={\mathbf {I}}_{{K_{I}}}\otimes {\mathbf{Q}} ^{\mathrm H}{\bm{\Phi}}^{\mathrm H} {\bm{\Phi}}{\mathbf{Q}},
\end{flalign}
\vspace{-0.8cm}
\begin{flalign}
&\hat{P}_{\mathrm{RIS}}^{\mathrm{act}}=P_{\mathrm{RIS}}^{\mathrm{act}}-\delta_{r}^{2}\|{\bm{\Phi}}\|^{2},
\end{flalign}
\vspace{-0.6cm}
\begin{flalign}
&{\mathbf {D}}_{i}={\mathbf {I}}_{{K_{I}}}\otimes{\bf{h}}_{i}{\bf{h}}_{i}^{\mathrm H},
\end{flalign}
\vspace{-0.6cm}
\begin{flalign}
P_{{i}}^{'}=\frac{P_{{i}}}{\eta_{i}}-\delta_{r}^{2}\|{\bf{h}}_{ r,i}^{\mathrm H}{\bm{\Phi}}\|^{2}.
\end{flalign}

However, it is noted that the constraint in  \eqref{P_ER_1} is non-convex,
which makes Problem \eqref{question_2} still intractable.
We then approximate the constraint \eqref{P_ER_1} by its first-order Taylor expansion as
\vspace{-0.2cm}
\begin{align}
{\mathbf {W}}^{\mathrm H}{\mathbf {D}}_{i}{\mathbf {W}}\geqslant2{\mathcal{R}} \{{\mathbf {W}}^{\mathrm H}(t){\mathbf {D}}_{i}{\mathbf {W}}    \}-{\mathbf {W}}^{\mathrm H}(t){\mathbf {D}}_{i}{\mathbf {W}}(t),
\end{align}
where ${\mathbf {W}}^{\mathrm H}(t)$ is the beamforming matrix  at the $t$-th iteration.
Then, Problem \eqref{question_2} is written as
\vspace{-0.4cm}
\begin{subequations}\label{question_3}
\begin{align}
& \max_{{\mathbf {W}},{\mathbf {V}}}\quad {\mathcal{R}} \{ {\bf b}^{\mathrm H}{\mathbf {W}}\}\!- \! {\mathbf {W}}^{\mathrm H}\!{\mathbf {A}}_{1}\! {\mathbf {W}}\!-\! \sum\limits_{k=1}^{K_{I}}|\rho_{k}|^{2} \mathrm{{Tr}}\{ {\bf{g}}_{k}{\bf{g}}_{k}^{\mathrm H} {\mathbf{V}} \}       \\
&\quad\mbox{s.t.} \quad {\mathbf {W}}^{\mathrm H}{\mathbf {B}}{\mathbf {W}}  +  \mathrm{{Tr}}\{{\mathbf{Q}}^{\mathrm H}{\bm{\Phi}}^{\mathrm H}{\bm{\Phi}}{\mathbf{Q}}{\mathbf{V}}\}  \leqslant \hat{P}_{\mathrm{RIS}}^{\mathrm{act}},\\
 & \quad\quad\quad \|{\mathbf {W}}\|^{2}+\mathrm{{Tr}}\{{\mathbf {V}}\}\leqslant P_{\mathrm {BS}}^{\mathrm{act}},\\
 & \quad\quad\quad 2{\mathcal{R}} \{{\mathbf {W}}^{\mathrm H}(t){\mathbf {D}}_{i}\!{\mathbf {W}}    \}\!+\!\mathrm{{Tr}}\{ {\bf{h}}_{i}{\bf{h}}_{i}^{\mathrm H} {\mathbf{V}}\}\!\!\geqslant\!\!P_{{i}}^{''}\!\!, \nonumber\\
 & \quad\quad\quad i\!\in\!\{\!1,\!\cdots\!\!,K_{E}\!\}\!,\\
 & \quad\quad\quad {\mathbf {V}}\succeq 0,
\end{align}
\end{subequations}
where $P_{{i}}^{''}=P_{{i}}^{'}+{\mathbf {W}}^{\mathrm H}(t){\mathbf {D}}_{i}{\mathbf {W}}(t)$.
Problem \eqref{question_3} is a convex problem which can be solved by CVX tools \cite{cvx}.

\vspace{-0.3cm}
\subsection{Optimizing the Reflection Matrix   ${\bf {\Phi}}$ of the Active RIS}
Given $\{{\bm \rho}, {\bm{\tilde{\gamma}}}, \{{\bf w}_{k}\}, {\mathbf{V}}\}$,
we consider to optimize $ {\bm{\Phi}}$ in this subsection.
First, by assuming rank(${\mathbf {V}}$) = $r_{E}$,
we can express ${\mathbf {V}}$ as ${\mathbf {V}}=\sum\limits_{k=1}^{r_{E}} {\bf v}_{k}{\bf v}_{k}^{\mathrm H}$ based on the eigenvalue decomposition (EVD).
Then, we define $\tilde{\Phi}=[{a_{1}e^{j\phi_{1}},a_{2}e^{j\phi_{2}},\cdots,a_{L}e^{j\phi_{L}}}]^{\mathrm H}$.
By substituting the expressions of ${\bf g}_{k}$ and ${\bf h}_{k}$  into Problem \eqref{question_1} and removing the constant terms, Problem \eqref{question_1} is rewritten as
\vspace{-0.1cm}
\begin{subequations}\label{question_5}
\begin{align}
&\max_{{\tilde{{\Phi}}}}\quad   {\mathcal{R}} \{ {\tilde{{\Phi}}}^{\mathrm H}{\bf e}\}-  \tilde{{\Phi}}^{\mathrm H}{\mathbf {F}} \tilde{{\Phi}}              &\\
&\quad\mbox{s.t.} \quad \quad {{\tilde{{\Phi}}}}^{\mathrm H}{\mathbf {J}}{{\tilde{{\Phi}}}}\leqslant P_{\mathrm{RIS}}^{\mathrm{act}},\\
 & \quad\quad\quad\quad {{\tilde{{\Phi}}}}^{\mathrm H}{\mathbf {R}}_{i}{{\tilde{{\Phi}}}}\!+\!2{\mathcal{R}}\! \{{{\tilde{{\Phi}}}}^{\mathrm H}{\bf {r}}_{i}\}\!  \geqslant \!\tilde{P_{{i}}},i\!\in\!\{1,\cdots,K_{E}\},\label{P_ER_3}
\end{align}
\end{subequations}
where
\vspace{-0.5cm}
\begin{flalign}
&{\bf e}= 2\sum\limits_{k=1}^{K_{I}}\sqrt{\alpha_{k}{(1+\tilde{\gamma_{k}})}}{\mathrm {diag}}(\rho_{k}^{*}{\bf{g}}_{r,k}^{\mathrm H}) {\mathbf Q}{\bf w}_{k} \nonumber\\                      &\quad\quad-\sum\limits_{k=1}^{K_{I}}|\rho_{k}|^{2} \big( {\mathrm {diag}}({\bf{g}}_{r,k}^{\mathrm H}){\mathbf Q}\sum\limits_{i=1}^{K_{I}}  {\bf w}_{i}{\bf w}_{i} ^{\mathrm H}{\bf{g}}_{d,k}\nonumber\\
&\quad\quad+ {\mathrm {diag}}({\bf{g}}_{r,k}^{\mathrm H}){\mathbf Q}{\mathbf V}{\bf{g}}_{d,k}\big),
\end{flalign}
\vspace{-0.75cm}
\begin{flalign}
&{\mathbf {F}}=\sum\limits_{k=1}^{K_{I}}|\rho_{k}|^{2} \big({\mathrm {diag}}({\bf{g}}_{r,k}^{\mathrm H}){\mathbf Q}\sum\limits_{i=1}^{K_{I}}  {\bf w}_{i}{\bf w}_{i} ^{\mathrm H}{\mathbf Q}^{\mathrm H}{\mathrm {diag}}({\bf{g}}_{r,k})\nonumber\\
&\quad\quad+{\mathrm {diag}}({\bf{g}}_{r,k}^{\mathrm H}){\mathbf Q}{\mathbf V}{\mathbf Q}^{\mathrm H}{\mathrm {diag}}({\bf{g}}_{r,k})  \nonumber\\
&\quad\quad+\delta_{r}^{2} {\mathrm {diag}}({\bf{g}}_{r,k}^{\mathrm H}){\mathrm {diag}}({\bf{g}}_{r,k})\big),
\end{flalign}
\vspace{-0.75cm}
\begin{flalign}
&{\mathbf {J}}= \sum\limits_{k=1}^{K_{I}}  {\mathrm {diag}}({\mathbf Q}{\bf w}_{k}) {\mathrm {diag}}({\bf w}_{k}^{\mathrm H}{\mathbf Q}^{\mathrm H}) \nonumber\\ &\quad\quad+\sum\limits_{k=1}^{r_{E}}  {\mathrm {diag}}({\mathbf Q}{\bf v}_{k}) {\mathrm {diag}}({\bf v}_{k}^{\mathrm H}{\mathbf Q}^{\mathrm H}) + \delta_{r}^{2}{\mathbf {I}}_{L} ,
\end{flalign}
\vspace{-0.7cm}
\begin{flalign}
&{\mathbf {R}}_{i}=  {\mathrm {diag}}({\bf{h}}_{r,i}^{\mathrm H})  {\mathbf Q}\sum\limits_{k=1}^{K_{I}}  {\bf w}_{k}{\bf w}_{k} ^{\mathrm H}{\mathbf Q}^{\mathrm H} {\mathrm {diag}}({\bf{h}}_{r,i})                                                      \nonumber\\
&\quad\quad+ {\mathrm {diag}}({\bf{h}}_{r,i}^{\mathrm H})  {\mathbf Q}{\mathbf {V}}{\mathbf Q}^{\mathrm H} {\mathrm {diag}}({\bf{h}}_{r,i})\nonumber\\
& \quad\quad+\delta_{r}^{2} {\mathrm {diag}}({\bf{h}}_{r,i}^{\mathrm H}){\mathrm {diag}}({\bf{h}}_{r,i}),
\end{flalign}
\vspace{-1.38cm}
\begin{flalign}
&{\bf r}_{i}=  {\mathrm {diag}}({\bf{h}}_{r,i}^{\mathrm H}) {\mathbf Q}\sum\limits_{i=1}^{K_{I}}  {\bf w}_{i}{\bf w}_{i} ^{\mathrm H}{\bf{h}}_{d,i} \!  +\! {\mathrm {diag}}({\bf{h}}_{r,i}^{\mathrm H}) {\mathbf Q}{\mathbf {V}}{\bf{h}}_{d,i},   &
\end{flalign}
\vspace{-0.4cm}
\begin{flalign}
& \tilde{P_{{i}}}=\frac{P_{{i}}}{\eta_{i}}-{\mathbf {W}}^{\mathrm H}{\mathbf {D}}_{i}^{'}{\mathbf {W}}-{\mathbf {V}}^{\mathrm H}{\mathbf {E}}_{i}^{'}{\mathbf {V}},
\end{flalign}
\vspace{-0.5cm}
\begin{flalign}
&{\mathbf {D}}_{i}^{'}={\mathbf {I}}_{{K_{I}}}\otimes{\bf{h}}_{d,i}{\bf{h}}_{d,i}^{\mathrm H},
\end{flalign}
\vspace{-0.7cm}
\begin{flalign}
{\mathbf {E}}_{i}^{'}={\mathbf {I}}_{{K_{E}}}\otimes{\bf{h}}_{d,i}{\bf{h}}_{d,i}^{\mathrm H}.
\end{flalign}
\vspace{-0.5cm}

Problem \eqref{question_5} is still non-convex due to the non-convex constraint \eqref{P_ER_3}.
Thus, we transform the non-convex constraint \eqref{P_ER_3} by its first-order Taylor expansion,
and constraint \eqref{P_ER_3} is transformed as
\vspace{-0.2cm}
\begin{align}
{{\tilde{{\Phi}}}}^{\mathrm H}{\mathbf {R}}_{i}{{\tilde{{\Phi}}}}\geqslant2{\mathcal{R}} \{{{\tilde{{\Phi}}}}^{\mathrm H}{\mathbf {R}}_{i}{{\tilde{{\Phi}}}} (t)   \}-{{\tilde{{\Phi}}}}^{\mathrm H}(t){\mathbf {R}}_{i}{{\tilde{{\Phi}}}}(t),
\end{align}
where ${{\tilde{{\Phi}}}}(t)$ is the phase shift vector at the $t$-th iteration.
Thus, the  constraint in \eqref{P_ER_3} is rewritten as
\vspace{-0.2cm}
\begin{align}
2{\mathcal{R}} \left\{    {\tilde{\Phi}}^{\mathrm H}\left({\bf r}_{i}+{\mathbf {R}}_{i}  {\tilde{\Phi}}(t)\right)   \right\} \geqslant \tilde{P_{{i}}^{'}},i\in\{1,\cdots,K_{E}\},
\end{align}
where $\tilde{P_{{i}}^{'}}={\tilde P_{i}}+{{\tilde{{\Phi}}}}^{\mathrm H}(t){\mathbf {R}}_{i}{{\tilde{{\Phi}}}}(t)$.
Problem \eqref{question_1} is reformulated as
\vspace{-0.5cm}
\begin{subequations}\label{question_6}
\begin{align}
&\max_{{\tilde{{\Phi}}}}\quad   {\mathcal{R}} \{ {\tilde{{\Phi}}}^{\mathrm H}{\bf e}\}-  \tilde{{\Phi}}^{\mathrm H}{\mathbf {F}} \tilde{{\Phi}}              &\\
&\quad\mbox{s.t.} \quad  {{\tilde{{\Phi}}}}^{\mathrm H}{\mathbf {J}}{{\tilde{{\Phi}}}}\leqslant P_{\mathrm{RIS}}^{\mathrm{act}},\\
 & \quad\quad\quad 2{\mathcal{R}}\! \left\{    {\tilde{\Phi}}^{\mathrm H}\!\left({\bf r}_{i}\!+\!{\mathbf {R}}_{i}  {\tilde{\Phi}}(t)\!\right)   \right\}\!\!\geqslant \! \tilde{P_{{i}}^{'}},i\!\in\!\{1,\cdots,K_{E}\},
\end{align}
\end{subequations}
which is a quadratically constrained quadratic program (QCQP) problem and can be solved by CVX tools.

\begin{algorithm}[t] 
\caption{AO framework of solving Problem \eqref{question_1} } 
\begin{algorithmic}[1] 
\State {Initial iteration number $t=1$, maximum number of iterations $t_{\mathrm{max}}$, feasible ${\bf w}^{(1)}$, \!${\mathbf V}^{(1)}$, \!${\bm{\Phi}}^{(1)}\!$,
error tolerance $\varepsilon$ and calculate the value of $\sum\limits_{k=1}^{K_{I}}\alpha_{k}R_{k}^{(1)}$;}
\State {Update $\tilde{\bm{\gamma}}^{(t)}$ by \eqref{re_1};}
\State {Update ${\bm{\rho}}^{(t)}$ by \eqref{re_2};}
\State {Update ${\mathbf {W}}^{(t)},{\mathbf {V}}^{(t)}$ by solving \eqref{question_3};}
\State {Update ${\bm{\Phi}}^{(t)}$ by solving \eqref{question_6};}
\State {If  $ | \sum\limits_{k=1}^{K_{I}}\alpha_{k}R_{k}^{(t+1)}-\sum\limits_{k=1}^{K_{I}}\alpha_{k}R_{k}^{(t)} | / \sum\limits_{k=1}^{K_{I}}\alpha_{k}R_{k}^{(t+1)} <
\varepsilon $ or $t\geq t_{\mathrm{max}}$, terminate. Otherwise, set $t \leftarrow t+1 $ and go to step 2. }
\end{algorithmic}
\end{algorithm}

\vspace{-0.4cm}
\subsection{Algorithm  Complexity}
Finally, Problem \eqref{question_1} is solved by alternately solving Problem \eqref{question_3}
and Problem \eqref{question_6} until convergence.
We summarize the proposed AO framework of solving Problem \eqref{question_1} in Algorithm 1.
It is noted that the main computation to solve  Problem \eqref{question_1} lies in
 alternately solving Problem \eqref{question_3} and Problem \eqref{question_6}.
We use $I_{a}$, $I_{b}$ and $I$ to denote the numbers of iterations
 for the convergence of
Problem \eqref{question_3}, Problem \eqref{question_6} and Problem \eqref{question_1}, respectively.
Then, the overall computational complexity of solving Problem \eqref{question_1} can be approximated by
${\cal O}\left( I \left(I_{a}K_{I}^{2}N^{2} + I_{b} L^{2} \right) \right)$.

\begin{figure}
\vspace{-0.4cm}
\centering
\includegraphics[scale=0.54]{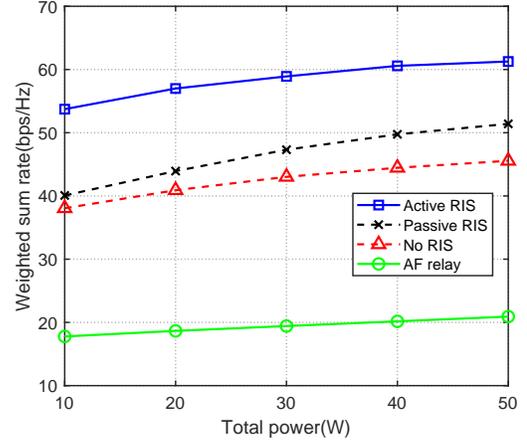}
\captionsetup[figure]{name={Fig.},labelsep=period}
\captionsetup{justification=raggedright}
\caption{ The WSR versus the total power when $N=4$ and $L=20$.}
\end{figure}

\vspace{-0.5cm}
\section{Simulation Results}
In this section, we provide numerical results to evaluate the performance of the active RIS-aided SWIPT system.
We assume that the BS and the active RIS are respectively located at (0 m, 0 m), (10 m, 10 m)
in a two-dimensional plane.
$K_{I}=4$ IRs are randomly distributed in a circle centered
at (30 m, 0 m) with a radius of 5 m, and $K_{E}=4$ ERs are randomly distributed in a circle centered
at (20 m, 0 m) with a radius of 5 m.
The large-scale fading of the channels are
modeled as $\small {{\rm{PL}}\! =\!-30\!  - \!10\alpha {\log _{10}}d} $ (dB), where $\alpha$ is the path loss exponent and $d$ is the link distance in meter.
In this work, we set {$\alpha = 2.3$ for ${\mathbf{Q}}$,
 $\alpha = 2.3$ for ${\bf{h}}_{ r,i}$,
  $\alpha = 2.5$ for ${\bf{g}}_{ r,k}$,
$\alpha = 3.2$ for  ${\bf{g}}_{ d,k}$,
and
$\alpha = 2.8$ for ${\bf{h}}_{ d,i}$.}
The small-scale fading  is assumed to be Rician distributed. 
For simplicity, the Rician factor is assumed to be 5.
The other parameters are set as follows: noise power of $\delta_{r}^{2}=\delta^{2}_{ \mathrm{IR}}=\delta^{2}_{\mathrm {ER}} =-80$ dBm,
{error tolerance of $\varepsilon = 10^{-3}$, minimum harvested power threshold of $P_{{i}} = 10^{-6}$ W.}

In order to illustrate the impact of the active RIS,
we compare the active RIS-aided multiuser SWIPT system with the following schemes:

$\bullet$
Passive RIS: It displays a passive RIS in the SWIPT system,
 which  means that only the phase shifts of the transmission signals are adjusted and
 there is no power amplifier at the RIS.

$\bullet$
No RIS: No RIS is to assist the SWIPT system,
which means that the BS only transmits signals to IRs and ERs through the direct links.

$\bullet$
AF relay: It displays an AF relay in the SWIPT system at the same location as the RIS
 in the SWIPT system.

{We adopt the power model in \cite{9734027,9377648}.
Thus, the power consumption models corresponding to the above schemes are given by
\vspace{-0.5cm}
\begin{align}
P_{{\mathrm {total}}}&=P_{\mathrm {BS}}^{\mathrm{act}} +P_{\mathrm{RIS}}^{\mathrm{act}}+L   (P_{\mathrm{C}}+P_{\mathrm{DC}}),\\
P_{{\mathrm {total}}}&=P_{\mathrm {BS}}^{\mathrm{pas}} +L   P_{\mathrm{C}},\\
P_{{\mathrm {total}}}&=P_{\mathrm {BS}}^{\mathrm{no}},\\
P_{{\mathrm {total}}}&=P_{\mathrm {BS}}^{\mathrm{af}}+ P_{\mathrm{relay}} +L   P_{\mathrm{T}} ,
\end{align}}
where $P_{\mathrm{C}}$ is the power consumption of the switch and
control circuit at each reflecting element,
 $P_{\mathrm{DC}}$ is the direct current biasing
power used by each active reflecting element,
{$P_{\mathrm{T}}$ is the dissipated
power at each antenna of the AF relay,
and $P_{\mathrm{relay}}$ is the transmit
power limit at the AF relay.}
{$P_{\mathrm {BS}}^{\mathrm{act}}, P_{\mathrm {BS}}^{\mathrm{pas}}, P_{\mathrm {BS}}^{\mathrm{no}},$ and $P_{\mathrm {BS}}^{\mathrm{af}}$ are
 the maximum transmit power of the BS in the corresponding schemes.          }
{Power consumption parameters of hardware devices are set as follows:
$P_{\mathrm{C}}= -10 $ dBm, $P_{\mathrm{DC}}=-5$ dBm, and $P_{\mathrm{T}}=10$ dBm.
We assume that  all schemes have the same total power $P_{{\mathrm {total}}}$,}
{and set $P_{\text{BS}}^{\text{act}}=P_{\mathrm{RIS}}^{\mathrm{act}}$,
$P_{\mathrm {BS}}^{\mathrm{af}}=P_{\mathrm{relay}}$.}

\begin{figure}
\vspace{-1.6cm}
\centering
\includegraphics[scale=0.44]{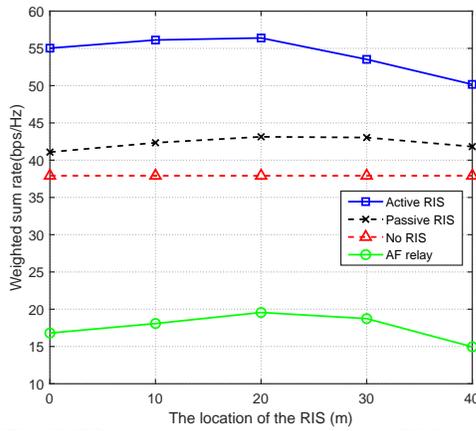}
\captionsetup[figure]{name={Fig.},labelsep=period}
\captionsetup{justification=raggedright}
\vspace{-0.3cm}
\caption{ The WSR versus the location of the RIS when $N=4$ and $L=20$.}
\end{figure}

Fig. 2 investigates the impact of the total system power on the WSR.
As seen from Fig. 2,
the WSRs of the scheme ``Active RIS'', the scheme ``Passive RIS'', the scheme ``No RIS'' and the scheme ``AF relay'' gradually increase with the
transmit power.
However, within the same total power consumption,
{it is observed that the scheme ``Active RIS'' achieves higher WSR than the scheme
``Passive RIS'', the scheme ``No RIS'' and the scheme ``AF relay'',}
which demonstrates that
displaying an RIS can enhance the WSR of  SWIPT systems and
active RIS can achieve best performance in an SWIPT system.

Fig. 3 depicts the WSR versus the location of the RIS.
We assume that the location of the RIS changes in the horizontal direction.
Comparing  ``Active RIS'' with ``Passive RIS'', ``No RIS'' and ``AF relay'',
it is observed that the WSR of the ``Active RIS'' is
always higher than the others.
In addition,
we can find that the WSR increases as the RIS is close to the location of the ERs,
and the WSR decreases as the location of the RIS is far away from the ERs,
which is
due to the fact that
the channel gain between the RIS and the ERs will become greater as the RIS's position approaches the ERs' positions,
and the ERs are easier to reach the thresholds of energy reception.
Thus, it leaves more space to jointly optimize the beamforming at the BS and reflection matrix
at the RIS to further improve the WSR.

Fig. 4 depicts the WSR versus  the number of RIS reflecting elements.
It is observed that the WSR of the scheme ``Active RIS''
increases slowly as the number of RIS reflecting elements increases,
when the number of RIS reflecting elements is small.
However, both the WSRs of the scheme ``Active RIS'' and the scheme ``Passive RIS'' decrease as the number of RIS reflecting elements
$L$ exceeds about 50,
which illustrates that the appropriate number of RIS reflecting elements
is able to make the system achieve the good performance,
and large number of RIS reflecting elements can cause performance loss, when the total power consumption is fixed.

\begin{figure}
\vspace{-1.6cm}
\centering
\includegraphics[scale=0.44]{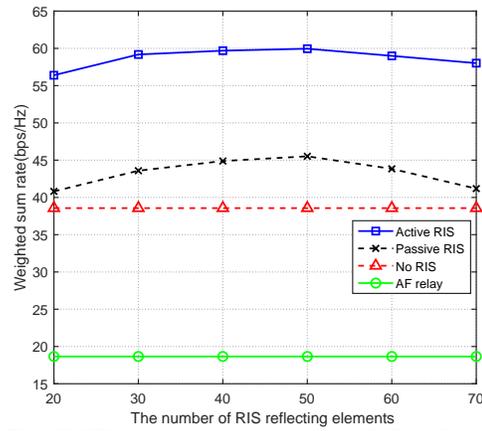}
\captionsetup[figure]{name={Fig.},labelsep=period}
\captionsetup{justification=raggedright}
\vspace{-0.3cm}
\caption{The WSR versus the number of RIS reflecting elements $L$ when $N=4$.}
\end{figure}

\vspace{-0.4cm}
\section{Conclusions}
This work studied an active RIS-aided SWIPT system.
We focused on maximizing the WSR of the IRs,
subject to the power requirements of all ERs,
transmit power limit at the BS and the amplification power budget at the RIS.
By adopting the FP
method and quadratic transform,
we transformed the original problem into a tractable form.
Then, the beamforming vector at the BS and the optimal reflection matrix at the RIS were
obtained  via the AO framework.
Under the same power consumption,
we compared the system gains for ``Active RIS'', ``Passive RIS'', ``No RIS'' and ``AF relay''.
Simulation results demonstrated that the active RIS-aided SWIPT system can achieve
better performance than the passive RIS/AF relay aided SWIPT system.

\bibliographystyle{IEEEtran}
\bibliography{IEEEabrv,myref}

\end{document}